# Smart City Governance in Developing Countries: A Systematic Literature Review

Si Ying Tan and Araz Taeihagh *

Lee Kuan Yew School of Public Policy, National University of Singapore, 469B Bukit Timah Road,
Li Ka Shing Building, Singapore 259771, Singapore; s.tan14@u.nus.edu
* Correspondence: spparaz@nus.edu.sg; Tel.: +65-6601-5254

**Abstract:** Smart cities that make broad use of digital technologies have been touted as possible solutions for the population pressures faced by many cities in developing countries and may help meet the rising demand for services and infrastructure. Nevertheless, the high financial cost involved in infrastructure maintenance, the substantial size of the informal economies, and various governance challenges are curtailing government idealism regarding smart cities. This review examines the state of smart city development in developing countries, which includes understanding the conceptualisations, motivations, and unique drivers behind (and barriers to) smarty city development. A total of 56 studies were identified from a systematic literature review from an initial pool of 3928 social sciences literature identified from two academic databases. Data were analysed using thematic synthesis and thematic analysis. The review found that technology-enabled smart cities in developing countries can only be realised when concurrent socioeconomic, human, legal, and regulatory reforms are instituted. Governments need to step up their efforts to fulfil the basic infrastructure needs of citizens, raise more revenue, construct clear regulatory frameworks to mitigate the technological risks involved, develop human capital, ensure digital inclusivity, and promote environmental sustainability. A supportive ecosystem that encourages citizen participation, nurtures start-ups, and promotes public–private partnerships needs to be created to realise their smart city vision.

## 1. Introduction

The smart city has emerged as a trendy political ideal that is widely adopted as a grandiose development vision in many countries. However, smart city development is not a novel or recent phenomenon. First surfacing in the mid-1800s in the new self-governing cities of the American west [1], the smart city became associated with the concept of 'smart growth', particularly in the context of transport, mobility, and planning in the 1990s [2]. Since the beginning of the new century, smart city development begin to garner serious interest among governments across the world [3]. The rise of smart cities at the turn of the 21st century can be explained by two perspectives. From the urban policy and planning perspective, the proliferation of smart cities is a result of both a technology push producing new levels of innovation capabilities and a demand-pull from cities seeking to address problems related to efficiency and sustainability [4]. From the economics perspective, the growth of smart cities in the context of the United States is hypothesised to be driven by a high concentration of human capital in these cities creating a pull factor for the in-migration of students to pursue high-quality education, as many major cities in the United States are also centres of higher education [5].

While the definitions and conceptualisations of what a smart city truly is remain highly heterogeneous, most studies agree that cutting-edge technological applications are some of the most salient components in smart city development for creating higher efficiencies in urban systems [1,6–13]. The incorporation of sensors and big data analytics enabled by the internet of things (IoT) is expected to promote operational efficiency in the design and administration of the cities that could bring about economic growth [6–8]. When cities are powered by IoT, they are expected to attract more economical investments by helping businesses differentiate and identify places to channel their capital [12]. The promise of IoT and of an interconnected urban system is fuelled by an unprecedented growth in information and communication technology (ICT) capabilities. This results in the emergence of new data processing technologies and processing logics such as quantum computing, machine learning, and deep learning; wider data availability; cheaper data processing power; and high levels of mobile technology penetration among the citizens, which could directly loop them into the decision-making processes [11]. In a nutshell, a smart city is perceived to present innovative solutions to a multitude of socioeconomic and environmental challenges faced by cities [1,9,10]. While government- and private sector–led technological solutions are brought to the front-and-centre of smart cities, a participatory approach has been equally emphasised that highlights the importance of citizens as co-creators and collaborators in the public value creation behind smart city governance and decision-making processes [9,10,12,14–16]. Likewise, including the citizenry in the development of smart cities and cross-sector collaborations involving many levels of actors, including citizens, are some of the important characteristics of smart city governance [9,10]. Fundamentally, citizens are core actors in society, and smart city development ought to reflect consistently on the extent to which ordinary citizens could benefit from a movement towards incorporating intelligence in various modalities of city operations [12].

To date, multiple case studies with the over-arching aim of developing theoretical frameworks that are centred on smart city governance have been written and published, especially among the developed economies. These empirical studies range from those discussing the implementation experiences, policy mechanisms, and transformation processes of smart cities [2,10,17,18] to those highlighting the drivers behind and barriers to smart cities [10], via those proposing new metrics for performance assessment, sustainability assessment, and benchmarking for smart cities [19,20]. There is also a growing concern with the ability of smart city development to safeguard environmental sustainability [2,7,12,18,20]. Besides, a good number of reviews have examined the state-of-the-art in smart city development and/or governance [1,6–8,14,16,21–23]. Some of these reviews focus on the applications of big data and artificial intelligence in smart cities and their potential to promote environmental sustainability [6–8], as well as the technological risks involved in smart city development consequent to the applications of IoT [21,22]. Others focus more on the governance aspect of smart cities [1,14,16,23]. The latter reviews in particular have attempted to identify the drivers behind smart cities [1], performance assessment frameworks for smart cities [14], the different components of smart cities [16,23], the desired outcomes of smart cities [1,23], and the contextual factors that influence both the components and the desired outcomes [23]. These comprehensive reviews, which explain how smart cities are planned, envisioned, and implemented in developed cities, have laid the foundation and provided the impetus the objective of this review, which explicitly focuses on consolidating evidence on smart city development in developing countries. We are specifically interested in understanding the state of smart city development in developing countries at this point, the unique sets of driving factors that propel the rise of smart cities in developing countries, and the exclusive challenges that developing countries face in the process of embarking on different smart city initiatives.

The smart city remains a largely opaque concept in developed countries [12,24,25]. Many studies differ in their emphasis on the various components of smart cities, the speed and nature of their governance processes, and the diverse legitimacy claims for smart city governance [16]. These suggest that the definitions and conceptualisations of smart cities in developing countries are likely to be murkier and could benefit from a systematic knowledge consolidation. Besides, the collision of the digital world with non-digital issues such as social justice, politics, ideology, legality, and regulation implies that the governance of smart city development is imbued with many layers of complexities [26]. The governance processes of smart cities could be even more intricate for some developing countries that are still en route to meeting the basic needs of all citizens. The complexity of the governance issues faced by developing countries, which are different from developed countries, suggests that a systematic literature review to find out what entails in the existing literature on smart cities development in developing countries is of paramount importance. Furthermore, rising rural-to-urban migrations will create more ecological, social, and infrastructural pressures in many major cities of developing countries. As of 2018, approximately 55% of the total population of the world reside in urban areas, and this trend is expected to rise consistently over the next few decades. By 2030, most of the 43 mega-cities (cities with 10 million inhabitants or above) are expected to be in developing countries [27]. Nevertheless, the high financial cost involved in infrastructure maintenance and the substantial size of the informal economies in developing countries pose unique challenges to the governments' smart city ideals which need to be unpacked [6]. Besides, privileging technology as the core of smart city development without aligning it with public values or understanding the extent of citizens' basic needs would transform smart cities into mere 'white elephants' [12]. Given the need to balance social needs with technology development in developing countries, improving the understanding of how smart cities can be governed in the face of competing developmental challenges becomes imperative.

This review aims to integrate the existing interdisciplinary literature depicting the experience of smart cities in developing countries to tease out the various aspects of smart city development and to inform smart city governance. To achieve this research aim, we undertake a systematic literature review of smart cities development in developing countries and pose the following research questions: (i) How are smart cities conceptualised and defined in developing countries? (ii) Why do government in developing countries embark on smart city initiatives and what are their specific goals? (iii) What are the specific drivers behind (and barriers to) smart city development in developing countries? The review found that technology-enabled smart cities in developing countries can only be realised when concurrent socioeconomic, human, legal, and regulatory reforms are instituted. Governments need to step up their efforts to fulfil the basic infrastructure needs of citizens, raise more revenue, construct clear regulatory frameworks to mitigate the technological risks involved, develop human capital, ensure digital inclusivity, and promote environmental sustainability. A supportive ecosystem that encourages citizen participation, nurtures start-ups, and promotes public–private partnerships needs to be created to realise their smart city vision. Through a process of systematic knowledge consolidation, this review contributes policy insights and practice recommendations that governments in developing countries could anchor on and draw lessons from in smart cities development.

## 2. Methods

### 2.1. Search Strategy

We undertook a systematic literature review to collate and synthesise evidence on the implementations and experiences of smart city development in developing countries. Two major academic databases that are the most widely used repositories to retrieve social sciences literature (Scopus and Web of Sciences) were utilised in the evidence-gathering process. Two comprehensive search strings—one specifying all the relevant keywords for smart cities and their analogous concepts, the other specifying the relevant keywords and conceptual equivalents for developing countries—were developed by the authors and combined using the parentheses 'AND' in the evidence search process (Table 1).

**Table 1.** Keywords included in the search strings.

| | |
|---|---|
| Smart cities and analogous concepts | (TITLE-ABS-KEY("smart cities" OR "smart city" OR "smart-city" OR "smart-cities" OR "Sustainable city" OR "Sustainable cities" OR "Sustainable urban development" OR "Sustainable urban developments" OR "Digital city" OR "Digital cities" OR "Eco City" OR "Eco cities" OR "Green city" OR "Green Cities" OR "Low Carbon City" OR "Low carbon cities" OR "Knowledge City" OR "Knowledge cities" OR "Resilient city" OR "Resilient cities" OR "Intelligent city" OR "Intelligent cities" OR "Liveable City" OR "Liveable cities" OR "Information city" OR "Information cities")) |
| Developing countries and conceptual equivalents | ALL(("developing countries" OR "developing country" OR "developing society" OR "developing societies" OR "middle income countries" OR "middle income country" OR "low income countries" OR "low income country" OR "lower middle income countries" OR "lower middle income country" OR "higher middle income country" OR "higher middle income country" OR "low and middle income country" OR "low and middle income countries" OR "less developed country" OR "less developed countries" OR "less economically developed country" OR "less economically developed countries" OR "underdeveloped country" OR "underdeveloped countries" OR "emerging market" or "emerging markets" OR "emerging economy" OR emerging "economies" OR "less industrialized" OR "less industrialised" OR "none industrialized" OR "none industrialised")) |

*2.2. Inclusion and Exclusion Criteria*

Five inclusion criteria were developed to guide the authors in gathering the relevant studies for this review. First, only studies published from 2009 to 2019 have been included. This is because the policy agenda on smart city development have only taken root in most developing countries in the past decade [28,29]. India, for instance, launched the 100 smart cities mission in 2015 when President Modi was newly elected [28], while the first policy paper on National Pilot Smart Cities in China was rolled out in 2012 [29]. Second, all review articles and empirical studies with a design ranging from quantitative or qualitative case studies, qualitative policy analysis and evaluation, conceptual studies, book chapters, and conference proceedings have been included in the search process. While it is often known that book chapters and conference proceedings are academic sources that are less rigorous than journal articles, we have decided to include them in view of the nascent development of smart cities in developing countries at this point. Third, we have only included studies comprising explicit information on the implementation, development, applications, dimensions, and goals of, as well as the drivers behind and the barriers/challenges to, the smart city in the developing country context. Fourth, we have also included studies that discuss analogous concepts of smart cities and their applications such as digitalisation, green solutions, and sustainability, as long as these analogous concepts and applications are discussed within the broader context of smart city development in developing countries. Fifth, while we have included studies with an over-arching focus on the public policy processes of smart cities, we have also included studies that discuss specific instruments utilised to advance the smart city agenda.

We also included three exclusion criteria to strengthen the evidence search process when deciding on the relevance of each article screened. First, we have excluded all studies that solely discuss smart city development in developed countries. Second, studies that discuss analogous concepts of the smart city without explicitly relating these discussions to smart city development in developing countries have been excluded. Third, we also have excluded empirical studies examining the impact of specific technology instruments or policies on certain environmental outcomes without situating the discussion within the broader context of smart city implementation and development within specific jurisdictions.

*2.3. Data Collection, Data Extraction, and Data Analysis*

All studies found through the initial search process using the search strings were incorporated in a spreadsheet. First, duplicates were removed. Subsequently, guided by the above inclusion and exclusion criteria, the remaining studies were screened for relevance based on their titles and information derived from the abstracts. After the initial screening process, the full text of all studies fulfilling all the inclusion and exclusion criteria were downloaded. The content of these texts was further examined to ascertain the extent to which they fit the scopes of this review.

A data extraction framework specifying details such as the title of the study, authors, methods, objectives, definitions of smart city, dimensions of smart city concepts, goals of the smart city, and drivers behind and barriers to smart city development were constructed to extract relevant information from each of the studies included this review. All data extracted were analysed by adopting a thematic synthesis approach that enables flexibility of analytical orientation, combining both deductive archetypes identified from earlier studies, and inductive sensitivity in teasing out new interpretive elements from new studies [30]. Following the three research questions that we posed, we identified different chunks of text in each study that exemplified the definitions/conceptualisations, goals, purposes, enablers, facilitators, barriers, and challenges of smart cities. This approach was further substantiated by line-by-line coding using thematic analysis to generate meanings from diverse sources and to unravel the conceptual nuances within the materials [31]. This approach helped discover patterns and relationships in the data, which subsequently enabled the identification of major themes and sub-themes. Such thematic analysis enabled us to identify relevant examples that aptly illustrated each of the themes and sub-themes.

Throughout the entire research process, the authors held multiple discussion to develop the search strategies and to design the research protocol. Consensus was achieved regarding the relevance of the studies included in the review, the scope of the review, and the extraction and analysis of the data.

**3. Results: Smart City Conceptualisations and Motivations in Developing Countries**

*3.1. Study Contexts and Characteristics*

The search process using the two categories of keywords above generated a total of 3928 hits. Guided by a set of inclusion and exclusion criteria, we identified a total of 56 studies relevant to this review. These 56 studies illustrated the policy processes and development experiences of smart city initiatives from a total of 12 individual countries across three continents—Asia, Africa, and Eastern Europe. Table 2 summarises the geographical location, level, and the design of the studies.

The characteristics of all the included 56 studies broadly fall into three categories: (i) studies that focus on in-depth analysis of smart city development process or policy process in a particular jurisdiction [28,32–66], (ii) surveys of citizens and experts regarding their views/opinions/preferences on smart cities adoption [3,67–73], and (iii) theorising/conceptualising the developmental trends of smart cities in the context of developing countries [74,75]. While most studies focus on the macro-policy implementation of smart cities, some studies dive into explaining the process of adopting a particular smart systems such as the incorporation of digital platform or ICT into public services [33,55,58,70], adoption of intelligent transport systems [61,76–78], and the adoption of IoT in public services [39,67]. Appendix A provides more detailed information about the 56 studies including country/initiative studied, method of the study and the aims/objectives of the study.

**Table 2.** A summary of the study contexts and characteristics[1].

| | |
|---|---|
| Geographical locations (country/continent) of the included studies by descending order (based on the number of studies identified) | India [28,32–44,67–70,79] |
| | China [45–53,71,76,80] |
| | Indonesia [54–57] |
| | Brazil [3,58,72] |
| | Malaysia [59–61] |
| | Vietnam [62,73] |
| | Mexico [63] |
| | Turkey [78] |
| | Egypt [81] |
| | Romania [64] |
| | Nepal [65] |
| | Ghana [66] |
| | Africa [74,75] |
| Study design by descending order (based on the number of studies identified) | Qualitative city-level case studies [32–34,37,38,42,44,45,47–49,50,52,53,56–58,60,62,65,76,80] |
| | Qualitative country-level case studies [28,33,35,38,43,44,47,48,52,55,57,58,60–62,65,66,[78]] |
| | Qualitative continent-level case studies [74,75] |
| | Quantitative surveys [3,67–73] |

[1] Contexts and characteristics of the 56 studies included in the review; please see Appendix A for more details.

*3.2. Conceptualisations of the Smart City in the Literature*

Among the 56 studies synthesised for this review, 14 attempted to conceptualise a smart city in the context of developing countries. It is almost unanimously held that a technology-enabled digital service is one of the key characteristics of a smart city [3,37,39,47,50,51,56,57,63,69,72,74,81,82]. For example, smart city development can only be realised "by the extensive application of ICTs in the urban management system" [57] (p. 62). These studies conceptualise the anatomical features of a smart city as a smart urban system that comprises four layers with different physiological functions: the perceptive layer (akin to human sensing organs), such as sensors, smartphones, smart cameras, and signal lamps responsible for collecting information; the network layer, which comprises the internet, IoT, and mobile communication network technologies, which operates like the human's nervous system in facilitating the transfer and storing of information; the platform layer, which is analogous to the human brain and is capable of analysing data in real time at the data centre or using cloud computing platforms; and the behavioural layer, which mimics human behaviour and enables policy-makers and administrators to make decisions based on the information collected and analyses conducted by the other three layers to strengthen the management of a smart city [51]. The ability to utilise ICT to collect and analyse data is at the heart of smart city development, as these insights help maximise resources in public service provision more effectively and efficiently [56].

Some studies do not discuss the smart city as a concept centred on technological capabilities, but discuss other peripheral concepts such as low-carbon cities [37,47] and eco-cities [50] in the context of smart cities. This chimes with many other studies that discuss the importance of ensuring that sustainability is upheld concurrently with smart city development, in that for a city to be smart, it also needs to be sustainable [3,37,47,50,56,57,72]. For a smart city's visions to be realised, it should include the sustainable elements of an eco-city, and it needs to be designed with the clear objective of elevating public services, improving government efficiency, and minimising negative environmental impacts through the creation of waste and pollution [50]. Similarly, smart city construction needs to be understood broadly as the integration of systems and services that could effectively reduce carbon emissions to develop a low-carbon economy that could sustainably fuel other developments to create cities that are smart, liveable, efficient, and affordable [37]. Combining elements of intelligence and sustainability is also perceived as imperative in smart city development to ensure that both physical and social environments are upgraded in tandem to improve the quality of life of the citizens [3,37,72].

It is also emphasised that the technology-enabled smart city cannot be realised without giving sufficient attention to all other aspects of development—including social development, economic development, legal and regulatory reforms, and human development—as these aspects form the concurrent conditions that smart city initiatives in developing countries have to meet [63,81,82]. For example, a study conducted in Mexico contends that, instead of solely depending on strengthening technology-driven automated procedures, smart city development is a sophisticated process that warrants "process redesign, political support negotiation, and the transformation of organisational structures and institutions, among other important factors" [57] (p. 62). Correspondingly, another study opines that economic development, social development, ecological development, and having the right political mechanisms are key characteristics that are important in defining a smart city [81]. These processes and mechanisms are highly intertwined and mutually interdependent to propel the success of smart city development [63]. Considerable attention should also be given to human development that entails the development of knowledge, skills, and participation, as without humans directing the translation of knowledge and the flow of ideas, the implementation of a smart city will not be realised [82]. Besides, smart institutions and laws are legal and regulatory conditions that are important in building a people-centric and inclusive smart city [74].

*3.3. Motivations for Smart City Development in Developing Countries*

This review found four over-arching objectives that propel governments in developing countries to prioritise smart city development as their public agenda.

1. Improving government efficacy in public service delivery: The foremost purpose of developing a smart city is to improve the government's efficiency in public service provision [38–40,50,58,66]. By capitalising on the technological prowess of IoT platforms, it is hoped that smart city development will improve public service efficiency by reducing transaction costs in the delivery of public services [40,74], facilitating a more efficient information structure to connect with citizens more effectively [50,69], improving production efficiency of firms [44], and inferring knowledge and insights generated from data collected from the IoT system to improve forecasting demand, quality monitoring, and anomaly detection of essential public services [39].
2. Improving citizen quality of life: The improvement of public service delivery has direct ramifications regarding improving the quality of life of citizens [37,51,66], which is another major goal of smart city development in developing countries. Unlike developed economies, many developing countries are still grappling with many developmental issues and how to upgrade their citizens' quality of life. It is thus hoped that developing countries could leapfrog the developmental process by riding on the global bandwagon of smart city development that is sprawling across the world to help overcome some of the most pressing challenges that developing countries are now facing. For instance, the 'smart mobility initiative' launched as part of the smart city developmental project in Ghana prioritises the improvement of speed of travel and safety with the ultimate aim of reducing pollution, traffic congestion, and noise pollution, transferring costs that could negatively impact the quality of life of the citizens [66]. Raising quality of life under a smart city agenda is crucial for governments in developing countries, considering that they will need to race against time to tackle deteriorating infrastructure, depleting energy resources, and deficiencies in essential public services (especially healthcare, sanitation, and housing) in the face of demand expansion resulting from an exponential growth in urban populations over the coming decades [37].
3. Promoting inclusive governance: Smart city development is also rolled out to promote inclusivity and collaborative spirits among various stakeholders [58]. For instance, in Brazil, inclusive governance espoused in the name of smart city development aims at encouraging information sharing to optimise joint decision-making among various stakeholders such as governments, citizens, private corporations, and third-party governments. This collaborative spirit also enabled different government agencies to integrate their services to the citizens. Besides promoting greater awareness to support well-informed decisions and joint actions to address issues involved in smart city development, fostering a collaborative spirit is important to build trust and confidence among citizens [58].
4. Inclusion of vulnerable and disadvantaged populations: Smart city development in developing countries aims to promotes the inclusion of vulnerable and disadvantaged populations [33,74,82]. It is perceived that, beyond improving the quality of life of the citizens, smart city development should ultimately uplift underprivileged citizens in developing countries by enhancing their capabilities [82]. It is only by addressing entrenched and structural injustice in the city and by promoting greater equality, diversity, and democratic participation in urban life that a smart city agenda can flourish in developing countries [33]. As such, the adoption of technology and ICT in smart city development needs to account for the needs of disadvantaged populations, especially those from the informal sectors who have not benefited from the conditions that have fuelled development in developing countries in the past [74]. Essentially, smart city development in developing countries can only be considered successful when it integrates the fundamental needs of all populations and positively contributes to one or more of the 17 sustainable development goals (SDG) [82].

## 4. Drivers of Smart City Development in Developing Countries

The results from the systematic review suggest that eight important driving factors propel the rise of smart city development in developing countries. Besides technology development, these drivers emphasise the development of economic and financing capacity, as well as the strengthening of regulatory development, human capital, and citizen engagement. In addition, the involvement of the private sector (which tends to have an edge in technology and resources) in creating a supportive ecosystem that promotes innovation will further boost smart city development in developing countries (see Table 3 for summary).

### *4.1. Financing Capacity of the Government*

The financing capacity of the government is arguably one of the most pivotal engines of the smart city. Developing countries have a higher propensity to be budget constrained and thus require a wide range of conventional and innovative financing instruments to beef up capital investments in developing smart cities [35,36,53]. In India, land-based financing instruments (i.e., property tax, vacant land tax, development impact fees, and betterment charges), congestion charging instruments (i.e., fuel taxes, motor vehicle taxes, and highway tolls), and debt-financing have been proposed as the ideal mix for the government in financing smart cities [35]. Other innovative financing instruments proposed are crowdfunding, earmarking government funding, monetising the big data collected, carbon offset, and creating smart government bonds [36]. China's experience in developing the International Low-Carbon City in Shenzhen has also seen the deployment of a wide range of innovative instruments [53]. Other than the municipal-owned Urban Investment and Financing Platform (which raises money through bank loans, short-term financing bonds, and registered private placement bonds), the city has also embarked on participatory and innovative financing approaches such as 'planning the village area as a whole' and 'the metro plus property development approach'. In the 'planning the village area as a whole' approach, residents in the local communities are included in the redevelopment of lands for high-end industrial purposes alongside the real-estate companies and industries. They are also opted as shareholders in the redevelopment package deal and benefit from the redevelopment process. In contrast, in the 'metro and property development approach', three major parties—the local government, the subway company, and real-estate developers—are involved. The local government offers the subway company the franchise to construct and operate the metro. The subway company then subcontracts the real-estate development rights to the real-estate developers on a profit-sharing model. This profit-sharing model enables the local government and the subway company to share land premiums and operating income generated from real-estate development. The subway company could, in turn, use some of these profits to invest in subway construction and maintenance, which will add value to real-estate development [53].

*4.2. Building a Strong Regulatory Environment that Fosters the Confidence and Trust of Citizens and Investors*

Building a strong regulatory environment that fosters the confidence of investors and the trust of citizens is another key driver of smart city development in developing countries [63,69,70,73,77,83]. To realise this, the government needs to take active steps to lay out clear sets of laws, regulations, and policies that guide smart city development [63,83]. These processes need to be transparent so that the government's credibility can be sustained among citizens and investors [73]. Increased transparency is a conduit towards governance improvement by which the government can generate better solutions by using a crowd-sourcing mechanism [73]. Besides, having strong and transparent regulatory institutions will enable the government to build the trust of citizens [69,70,73] and the confidence of investors [73]. Some of the fundamental issues that government has to confront in the development of the smart city are the privacy and security issues stemming from the use of IT-enabled services and the massive amount of data collected by IoT [69,70,77]. To build trust, the government needs to enact privacy laws that respond to the concerns of citizens and investors on data privacy and security breaches. In particular, clear institutional arrangements for data collection and data sharing integrated from different sources are important [77]. These arrangements should spell out the details of ethical standards and safety procedures corresponding to the various privacy and security risks. For instance, important issues to address include stipulating who has the rights to data ownership and data access, which party is liable for data breaching, and who should benefit from the profits generated by data [77].

*4.3. Technology and Infrastructure Readiness*

Technology and infrastructure readiness will help lay the physical foundations for all smart city initiatives and are important catalysts for smart city development in developing countries [38,45,48,62,75,81–83]. The availability of high-quality, high-security, and privacy-preserving wireless infrastructure and service-oriented information systems that share public service information with the citizens will improve operational and administrative efficiencies [48,62,72,75,82]. These information systems and service platforms would enable citizens to perform various transactions or queries online [72,82]. Besides, the government can tap into these systems and platforms to enhance public services and resource management by setting-up an integrative e-government portal [45,81,82]. For example, as part of the quest to create 'knowledge cities' in Romania, the Romanian government launched a series of e-government portals including an e-Procurement system for public acquisitions, a virtual payment desk for the electronic attribution of international authorisations and transport goods, a computer-assisted learning system for the citizens, an e-payment system for taxes, and an IT-based system centralising the announcement of high school examination results and the allotment of university admissions [81].

To accrue the benefits technology could bring to both governments and citizens in smart city development, an advanced IoT system with functional physical infrastructure (such as smart cameras, sensors, and actuators that collect data) [38] and a centralised analytics system capable of processing, analyzing [75], and inferring knowledge from data and transferring this information to the necessary hubs [38] are required. For this integrated network to function effectively, regular maintenance of both the physical and network infrastructure must be conducted [83].

*4.4. Human Capital*

Human capital is an indispensable driver in smart city implementation in developing countries [44,70,77,82]. Having citizens that are sufficiently educated and technically competent in navigating the smart city environment, including knowing the opportunities that different smart city initiatives are accruing, is highly important for the development of smart cities [82]. Beyond this, strengthening the technical capacity of human resources in terms of the knowledge, expertise, and motivation needed to implement IT-enabled services would enable governments to roll out various smart city initiatives expeditiously [70,77]. Operational efficiency will be enhanced if the IT staff are adept in handling security and privacy issues embedded in all IT-enabled services [70]. In addition to handling risks, technical capacity is also required in other aspects of operations within the IoT system, such as data collection, data integration, data management, data analysis, and information service provision [77]. In developing countries, many cities do not have the technical capacity to perform data integration by developing their own applications or using data aggregators such as Application Programming Interfaces (APIs) [77]. To overcome this problem, city administrators could focus on publishing good data and outsource the management of APIs to third-party developers [77].

*4.5. Stability in Economic Development*

Economic stability is another key driver of smart city development in developing countries [71,83]. This factor is a particularly important consideration for foreign investors evaluating their financial risks and the likelihood of a smart city project's success [71]. Based on a quantitative analysis of the drivers of smart cities, a city's GDP growth (for instance) has a direct impact on bringing more financial capital for hardware development, characterised by more investment in transportation infrastructure and telecommunication and attracting human capital inclined towards leading a better quality of life [24]. Smart city development will facilitate the flow of capital, and it is important that economic growth in the smart city remains constant to attract capital investment [83].

*4.6. Active Citizen Engagement and Participation*

Active citizen engagement and participation is another key driver of smart city development in developing countries [42,49,58,63,69,71,82]. Active citizen engagement needs to embrace the notions of citizen empowerment [58,82], digital inclusion [63,82], collaborative governance [3], behavioural change to ensure that the ultimate objectives of smart city development are realised [71], and a benefits-sharing approach in ensuring that citizens have a fair share in enjoying the fruits of smart city development [49].

Citizen empowerment entails building a strong sense of awareness and ownership among the citizens, encouraging inter-sectoral relationships that include the citizens, and promoting active participation in providing feedback at every stage of the policy process [58,82]. Digital inclusion is a process whereby smart city strategies are designed with clear target populations or communities, including identifying the specific skills, limitations, and deviations that these communities can bring to the smart city project [63]. During the implementation of smart city initiatives, citizens need to be made aware of the benefits connected to various smart city tools and given the opportunity to provide feedback to the government for improving the tools [82]. Collaborative governance is another citizen-centric process that comprises three key actors—governments or policy makers, citizens, and researchers—that collectively work toward achieving smart city goals [3]. It is a highly participatory process underscored by the interaction between policy, community, and research in making decisions that improve citizens' quality of life [3]. Behavioural change is especially pertinent in the context of eco-city development, which intends to promote low-carbon lifestyles among the citizens. For instance, in the case of China, it was acknowledged that the awareness and behaviours of citizens would need to alter dramatically to achieve the cities' targets in promoting energy efficiency and environmental sustainability [71]. A benefits-sharing approach in smart city development upholds citizen welfare in smart city development. For example, in Shenzhen in China, villagers were given financial stakes in smart city development to allow them to benefit from a portion of the land sale for commercial use, and their land titles and profits from land development were protected by legislation [49].

*4.7. Knowledge Transfer and Participation from the Private Sector*

While the government's steering capability is crucial in smart city development, knowledge transfer and participation from the private sector would help to boost the probability of success [44,49,71,77]. Private sector participation is important as it is seen as complementing three areas of the government's roles in smart city projects: financing, mentoring, and acting as incubators for testing new ideas [44]. Besides supporting the government's role as a standalone entity, private sector participation can occur in the form of effective public–private partnerships when implementing smart city projects [77]. For instance, China is one of the developing countries that has embraced private sector participation, both locally and internationally, in various smart city projects [49,71]. Some of the most high-profile smart city projects in China, such as the ones in Guangzhou and Shenzhen, are primarily driven by multi-actor partnerships between local governments, state-owned developers, and the international private companies that comprise architecture firms, consultancy firms and other international developers [49]. In supporting smart eco-city development in China, knowledge transfer from the private sector in areas such as clean energy development and the energy and carbon trading markets not only helps promote the adoption of low-carbon technology and the construction of green building among the developers; it also reduces the burden of government in financing these capital-intensive projects [71].

*4.8. Creating a Supportive Ecosystem that Promotes Innovation and Learning*

Creating a supportive ecosystem that promotes innovation and learning also helps drive smart city development in developing countries. This mechanism can be created either through a government-led incubator system [44,77] or through an experimental approach in policy implementation [47]. For instance, both the state and regional governments in India have been very active in creating start-up incubators to promote citizen-led IoT interventions [44]. A technical ecosystem focusing on the cultivation of knowledge exchange and learning, as well as collaborations through workshops, fora, hackathons, and conferences, can also be set up to combine experiences and skills sets from a range of different actors including technology vendors, manufacturers, developers, system integrators, solution providers, data analysts, network designers, cybersecurity specialists, investors, and entrepreneurs [77]. In China, a combination of top-down and bottom-up design from both the national government and city governments creates a nested structure that enables multiple layers of policy experiments on low-carbon transition to be conducted [47]. While the national government formulates experimental ideas and selects different cities to test various smart city policy measures, the local governments drive the implementation and evaluation of these experiments. These policy experiments at different levels have helped to shape policy practices and promote constant policy learnings for both the city and national government.

**Table 3.** Drivers for smart city development in developing countries.

| Drivers for Smart City Development in Developing Countries |
|---|
| • Financing capacity of the government [35,36,53]; |
| • Building a strong regulatory environment that fosters the confidence and trust of citizens and investors [63,69,70,73,77,83]; |
| • Technology and infrastructure readiness [39,46,63,64,76,77,82,83]; |
| • Human capital [44,70,77,82]; |
| • Stability in economic development [71,83]; |
| • Active citizen engagement and participation [42,49,58,63,69,71,83]; |
| • Knowledge transfer and participation from the private sector [44,49,71,77]; |
| • Creating a supportive ecosystem that promotes innovation and learning [44,47,77]. |

**5. Barriers to Smart City Development in Developing Countries**

Smart city development in developing countries faces exclusive challenges that are fundamentally different from those facing developed countries due to existing socioeconomic challenges. This systematic review has identified 10 unique social, economic, environmental, political, and regulatory barriers to smart city development in developing countries that need to take precedence before smart city visions can be realised in developing countries. These barriers are highlighted below and summarised in Table 4.

*5.1. Budget Constraints and Financing Issues*

Budget constraints and financing issues are among the most significant barriers identified in smart city development in developing countries [40,41,43,48,67,74,78,81]. Budget constraints restrain the ability of governments to develop smart cities, even though smart cities are projected to be more cost-effective in the longer term [40]. In some countries with multiple levels of jurisdiction on fiscal resource allocation, over-dependency on state and central governments for developmental funds also hampers the ability of local governments to generate their own revenues through taxes and levies to finance a mega-project like a smart city [43]. Besides, the imperative of addressing other developmental challenges, such as the poverty and inequality witnessed through the rise of slums in some cities in Egypt [81] and pervasive youth employment in Africa [74], pose as competing developmental priorities to the government when engaging in smart city projects. These developmental issues perpetuate urban poverty and are directly connected to other social issues in the cities, such as high levels of crime and the rise of the informal economy dominated by low-level employment or unskilled jobs. A labour market dominated by an informal economy with a limited knowledge base for smart city development has hindered the government's ability to expand formal employment, which requires higher levels of technological skills sets to drive smart city development [74].

The financial viability of the government in financing smart city development is also halted by the astronomical costs involved for both governments and citizens [41,48,67]. For instance, in China, the Dong-tan eco-city project was estimated at around $1.3 billion USD, and this amount was unlikely to be shouldered by either business or government alone. This was compounded by uncertainties regarding the potential of this eco-city project to generate sustained profits. As a result, developers were not incentivised to invest their capital into the project, hindering the ability of governments to fast-track the development of some eco-city projects in China [48]. In India, the high cost of smart device ownership and the high electricity consumption resulting from the use of smart devices were found to deter smart city adoption as citizens felt the pinch when electricity bills increased [67], and it is uncertain if these costs are partially borne by other types of financing sources from the government.

*5.2. Lack of Investment in Basic Infrastructure*

Despite advocating for the smart city agenda, some developing countries are still held back by the lack of investment in basic infrastructure. Basic urban infrastructure, such as having proper water drainage and sewerage systems, are requisite for any city to thrive, but some cities still face a shortfall in providing these services [37,43,65]. In India, smart city adoption is held back by the lack of provision and maintenance in basic infrastructure, which result in uneven development and low-quality infrastructure in some cities, particularly in the slum areas [43]. In addition, the mounting population pressure is making infrastructure such as underground sewerage and water drainage systems unsustainable [37]. Under-provision of door-to-door garbage collection services and inefficient municipal solid waste management systems in some cities in South Asia are also hampering the effort of smart city development, as governments still struggle to meet the basic public service needs of their citizens [37,65].

*5.3. Lack of Technology-Related Infrastructure Readiness*

Many developing countries are falling behind developed economies in terms of technology-related infrastructure readiness, and this poses an imminent challenge to smart city adoption [33,51,66,84]. The level of technology-related infrastructure development is heterogeneous among developing countries, and the issues encountered are also different, ranging from lack of internet penetration [33] and lack of internet connection for business information exchange in Ghana [66] to the lack of local technical capacity to develop the core technologies needed to drive smart city development [51]. In large cities like Ahmedabad in India, the development of the smart city is constrained by the low internet penetration rate among the households and unequal access to digital infrastructure among the populations [33]. Likewise, in Ghana, costly internet connection services hinder the widespread adoption of Wi-Fi on privately owned commercial transport systems, which has slowed down information exchange for business owners, especially when travelling between cities [66]. In China, lack of local technical capacity in developing some of the core technologies needed in smart city development has become a concern, as this situation might predispose the country to potential technological risks that affect its national security and sovereignty. Dependence on foreign companies in core technologies such as various dynamic and spatial information systems, database management, and operation solutions potentially leads to security risks concerning the leak of confidential information [51]. These technology-related infrastructure gaps will ultimately result in a digital divide [33], affect business productivity, and expose cities to unintended risks [51].

*5.4. Fragmented Authority*

Fragmented authority across various public institutions also presents as a barrier to smart city development in developing countries [28,33,49,77,84]. As smart city development involves multiple stakeholders from a variety of public and private sectors, forming strong partnerships and having a central authority to steer the entire development process is crucial. The lack of unifying strategies for smart cities led by a central authority [33] and the lack of coordination and cooperation among a city's operational networks [77,84] inhibit the efforts of governments to roll out smart city projects swiftly and effectively. In India, many smart city projects have been launched independently by different agencies at the state and city levels, and these plans were often not coordinated by a central authority. This has resulted in a divergence of aims, the repetition of smart city plans, and the overlapping of responsibility in the execution of many smart city projects [33]. Likewise, a fragmented institutional mandate and the lack of coordination mechanisms across public service institutions has resulted in the lack of a centralised data controller platform to aggregate data for purposes of analysis. In particular, it has been found that more than one-third of the designated smart cities in India have not been able to collect and report data on various public services [28]. This fragmented authority could result in multiple complications in the governance of smart cities, including the deployment of ineffective policy instruments, operational inefficiency, the reluctance of local governments to concur to the development plans of smart cities, and the formulation of overambitious and unrealistic goals for smart cities [49].

*5.5. Lack of Governance Frameworks and Regulatory Safeguards for Smart Cities*

Lack of governance frameworks and regulatory safeguards is another major barrier identified in smart city development in developing countries [51,55,73–75,77,84]. While many developing countries are riding the bandwagon of smart city development for socioeconomic gain, there is essentially a lack of clear governance frameworks specifying the policy objectives, development strategies, regulatory norms, and evaluation models of smart city development in many cities in developing countries [55,73,84].

With respect to regulatory safeguards, the breach of data privacy and security surfaced as a major concern among citizens in different countries [51,75,77]. Integration of multiple networks in the cities and recent advancement in big data analytics indicate that many citizens could be subject to greater informational safety risks such as identity theft and other forms of cybersecurity attack [74]. Without clear regulatory safeguards in restricting data use, the privacy and security of a wealth of real-time data collected from terminal sensors through scanning, picturing, locating, and tracing could be endangered [51]. For example, smart mobility solutions such as ride-hailing and autonomous vehicles are continuously collecting data from users and could easily predispose users to the abuse of their privacy and data security through a malicious attack and many frauds designed on their mobile applications [77]. Even though many public service platforms are interconnected for better resource-sharing and more effective inter-system communication, this could expose the system to a higher risk of information leaking and data theft, potentially resulting in the collapse of the entire information system network [51]. In the implementation of satellite enhanced telemedicine and e-health in Africa, the governance of many privacy and confidentiality legal issues pertaining to the ownership of patient's data, security, and access to the clinical information systems by patients and care providers remained vague and unregulated [75]. Furthermore, the wiring-up of the ICT network means there is greater room for terrorist ideologies to be easily propagated in the cities if the governments do not devise clear regulatory and governance frameworks to monitor such illicit activities online [74].

*5.6. Lack of Skilled Human Capital*

Lack of skilled human capital, especially citizens that are adept in handling technology-enabled functional roles, is another major barrier in smart city development in developing countries [33,37,40,43,44,78,84]. In developing countries, the spectrum of skills sets lacking in smart city development includes the lack of formal skills in ICT [37], the lack of technology-related skills among the bureaucrats [40], and the lack of technical knowledge among planners [84]. In India, the government's capacity in smart city development has been hampered by a skills deficit, as the country has not been producing a sufficient number of planners evenly distributed across its various cities [43]. For example, in Bhubaneswar, a city located at the northwest of India, the Development Authority had only one qualified urban planner out of its total staff strength of 1137 personnel to oversee the planning functions of the entire city that were inhabited by close to one million population [33]. This situation demonstrates that most cities in India are ill-equipped to plan and manage urban systems comprehensively. The inability to recruit talent with relevant skills sets in second-tier cities is also forcing enterprises and smart city-based start-ups to migrate to other major cities for better prospects in recruiting higher-skilled talent [44]. In Turkey, increasing ICT human resources and competence in tertiary institutions, public agencies, and the private sector has been explicitly laid out as part of the country's smart city blueprint to address the deficit in skilled human capital in driving smart city development [78].

*5.7. Lack of Inclusivity*

Lack of inclusivity of the original residents, including the poor and the marginalised populations whose lands are often acquired for smart city projects, is another major challenge smart city development in developing countries needs to confront [34,41,45,74]. Forced land acquisition [34], forced displacement of original residents [45], and spatial segregation leading to the creation of class enclaves [41] are some of the documented scenarios that violate the principle of inclusivity in smart city development. In China, forced displacement of the original residents in Dong-tan eco-city disrupted the original social fabric and affected livelihoods. Furthermore, the original residents did not benefit from the development, as the new residential units that were built under the eco-city project were largely unaffordable to them and mostly ended up in the hands of speculative investors and upper-middle-class residents from nearby regions [45]. The plight of the original residents occupying the lands acquired for smart city development has also been observed in African countries, where poor land administration services have disabled the process of acquiring legal proof of land ownership among the residents, rendering them not only liable to eviction but also to exclusion from enjoying the economic gains acquired from the land development [74]. Similar circumstances have been observed in India, where forced land acquisitions from the poor tribal people and skyrocketing housing prices at levels unaffordable even to middle-class Indian citizens have drawn sharp criticism [34].

*5.8. Environmental Concerns*

One of the most challenging of the aspects governments have to grapple with in smart city development in developing countries is the various environmental issues that could ensue from the massive development that smart cities have brought [34,42,43,65,74,80,84]. These issues tend to result from the sudden ecological stress imposed on the environment due to the large migration from rural to urban areas to take advantage of the opportunities arising from smart city development. This phenomenon has resulted in many unplanned settlements in the cities and brings about congestion and pollution, besides other forms of resource allocation issues [42,65,74,84]. One of the most significant environmental concerns due to the sudden increase in urban population is the ecological pressure exerted on the environment from over-extraction of groundwater resources. Unregulated or illegal groundwater extraction reduces the depth of the groundwater table, and dumping of solid waste or industrial effluent carrying toxic chemical compounds into water bodies has also resulted in widespread environmental pollution [43]. Another major environmental issue concerned with the sudden rise in a city's population is a lack of capacity in the municipal government to manage household solid waste management. In Nepal, lack of human resources, lack of proper infrastructure (including vehicles to transport waste), and poor route planning has led to inefficient solid waste management services in Bharatpur metropolitan city. These have caused secondary adverse effects, such as improper dumping and open burning of the waste, which pose a threat to environmental sustainability [65]. Some African countries are also still plagued by a lack of the basic infrastructure to direct floods and mitigate droughts, and these problems are now aggravated by pollution and climate change, which causes the increased frequency of droughts, floods, and other forms of environmental stress. Without addressing these, it is predicted that scarcities in water and energy will ultimately lead to famine and disease, threatening the growth and sustainability of these countries [74]. Added to the multitude of environmental concerns stemming from smart city development is the disturbance to the equilibrium of the natural ecosystem. For instance, the massive development that takes place after the inception of an eco-city project in Dongtan City in China, is predicted to affect the entire ecosystem in that area, which may have been the wintering ground for some of the rarest bird species in the world and fertile spawning ground for salt marshes and several fish species [80].

*5.9. Lack of Citizen Participation*

Lack of citizen participation or public involvement presents another major challenge for governments in smart city development in developing countries [33,41,43,48,84]. Many smart city projects had insufficient or no public involvement in the preparation of metropolitan development plans, leading to a monolithic policy formulation process and fragmented decision-making [43,48,84]. Ambiguity in the policy process of smart city development, especially at the formulation and implementation stages, has also been demonstrated by a lack of clarity in establishing how citizens should be involved in the consultation and planning processes, which entailed determining the specific sociodemographic profiles of citizens that should be consulted and aspects of smart city design that should incorporate citizen participation [41]. In India, the low level of participation in the governmental platforms in the majority of the cities may be a direct consequence of the lack of local engagement by the central government, which took charge of managing and operating these governmental platforms [33].

*5.10. Technology Illiteracy and Knowledge Deficit among the Citizens*

Last but not least, technology illiteracy and knowledge deficit among the citizens could pose an obstacle to a government in realising its smart city vision in developing countries, especially those countries that are lower in the human development index [40,66,84]. The technology illiteracy of citizens can hinder the uptake of technology, the scaling up of technology adoption, and the realisation that positive network effects require a large user base comprising a high number of technology-literate citizens [40,66,84]. In a large developing country like India, which comprises a significant proportion of the population that is still illiterate, the paucity of knowledge in the use and application of the internet and modern technology such as the smart meter is hindering the government's effort to expand technology use among the citizens [40]. This situation is exacerbated by the lack of access to technology in some parts of the country [84]. Likewise, in Ghana—a country with an illiteracy rate as high as 25% among citizens aged 11 and above—the knowledge deficit in the use of technology and the slow rate of technology penetration has undermined the operational efficiency of smart city development and hampered the speed of information mobility to the citizens [66].

**Table 4.** Barriers to smart city development in developing countries.

| Barriers to Smart City Development in Developing Countries |
| --- |
| • Budget constraints and financing issues [40,41,43,48,67,74,78,81]; |
| • Lack of investment in basic infrastructure [37,43,65]; |
| • Lack of technology-related infrastructure readiness [33,51,66,84]; |
| • Fragmented authority [28,33,49,77,84]; |
| • Lack of governance frameworks and regulatory safeguards for smart cities [51,55,73–75,77,84]; |
| • Lack of skilled human capital [33,37,40,43,44,78,84]; |
| • Lack of inclusivity [34,41,45,74]; |
| • Environmental concerns [34,42,43,65,74,80,84]; |
| • Lack of citizen participation [33,41,43,48,84]; |
| • Technology illiteracy and knowledge deficit among the citizens [40,66,84]. |

## 6. Discussion

This review establishes fundamental differences in the characteristics of smart city governance between developed and developing countries. In particular, the 10 barriers to smart city development in developing countries identified from the review are specific developmental challenges that developed countries with higher levels of human development index have long resolved in their transformations from developing to developed countries in the past [85]. To realise smart cities' visions, it is important for countries or cities to harness the enablers that are prominent in their respective contexts. However, the different challenges facing smart city development described in this review may not apply to all cities in developing countries. For instance, the nature of the regulatory environment, presence of foreign direct investment in smart cities development and political structure of a country (top-down versus bottom-up administrations), are institutional factors that will affect the speed of adoption of smart cities' initiatives. In fact, developing countries have been categorised into five major groupings based on their respective socio-political and macroeconomic indicators such as poverty, income inequality, productivity, innovation, GDP, political freedom, overall governance, carbon emission, and external flow. These are important factors to be accounted for when ascertaining the developmental needs for smart city development in developing countries [86]. Some challenges may be more prominent in one context than in another, implying that priorities in addressing these challenges should be based on their respective salience and urgency to the local governments in different cities. Even though the differentiation of smart city governance in developing countries based on these factors is beyond the scopes and objectives of this review, it would be an important next step for the research on smart cities in developing countries to pursue.

This review has shown that the conceptualisation of smart cities in developing countries centres on the incorporation of digital technologies in the cities' infrastructure and services. Even though technology inclusion is one of the most important hallmarks of smart cities, the incorporation of digital technologies will only be meaningful if other concurrent conditions are actively pursued. These concurrent conditions include all aspects of socioeconomic development and regulatory reforms. To build a robust implementation structure, it is critical to use technological tools such as blockchain technology to provide a secure communication platform that minimises privacy and cybersecurity threats [87], as well as ensuring different information systems are well-integrated to enforce financial and administrative accountabilities [88]. Building smart cities within a designated period is fundamentally different from running and maintaining smart cities well, which require long-term planning, constant refinement, and revisions of public policies. To catch up with smart cities in developed countries, governments at the local and national levels in developing countries will need to bolster their efforts to strengthen some of the drivers of smart cities identified in this review.

In developing smart cities in developing countries, it is important to account for contextual conditions. Large disparities can be observed between cities even within the same country. Hence, planning and development need to account for the unique local characteristics of these cities depending on their developmental conditions such as their affluence, economic profile, stock of skilled human capital, technology literacy among the population, bureaucratic efficiency, and the overall policy capacity of the local government. In China, more developed mega-cities and provincial capital cities may benefit directly from integrated city planning that embodies most of the smart city characteristics found in major cities in developed countries; in contrast, in less developed second- or third-tier cities, smart city initiatives may choose to focus more on large-scale infrastructure upgrading [46]. The difference in infrastructure needs between cities in the same country is likely to hold true for many other developing countries.

Besides journal articles, this review draws upon book chapters and grey literature such as conference proceedings, government documents, research, and technical reports etc.) in the synthesis process. A fair amount of studies on smart city development and experiences in developing countries have been published as grey literature, and the lack of rigorous empirical research on smart cities development in developing countries is another limitation of this systematic review. Nevertheless, we are of the opinion that it is important to incorporate the grey literature in the review, as they are likely to represent the early scholarly voices in a nascent field that has not quite taken root in developing countries. This review also highlights a major gap in the literature on smart city development in developing countries, especially from the perspectives of urban planning and management, public administration, policy, and governance. This poses as a limitation for this review to examine the policy processes of smart cities in developing countries, land further hinders conceptual studies that aim to propose frameworks on the governance of smart cities in the context of developing countries. Lack of such in-depth policy analysis has inhibited nuanced understanding of the unique challenges faced by different developing countries which require targeted policy interventions. This review, hopefully, will be a starting point in the systematic creation of a broad understanding on the development and implementation challenges of smart cities in developing countries, and will inspire many empirical studies to be conducted in the future.

*6.1. Policy Implications for Smart City Governance in Developing Countries*

The findings on the drivers behind and barriers to smart city development present several policy implications for smart city governance in developing countries.

(i) Stepping up the effort to fulfil basic infrastructure needs

Some developing countries, especially those with lower levels of human development, are on a treacherous journey to build smart cities that could benefit all segments of the population. As some developing countries are still grappling with the need to meet the basic infrastructure requirements of the citizens, it is important to prioritise social development such as electrification, sanitation, clean water provision, and road construction before embarking on other ambitious developmental goals. While some of these challenges could be addressed with technology, it is important to note that technological advancement should not be fast-tracked to benefit a handful of elites and leave out larger segments of the population. To pursue healthy and multi-scalar development, cities should not overlook the provision of basic necessities such as water, energy, public health, sanitation, and transportation, and should ensure the equitable distributions of these necessities to all segments of the population [89]. It has been advocated even in developed cities that it is crucial for smart cities to incorporate the spirit of inclusive participation by placing citizens at the centre of all policy actions and decisions [90].

(ii) Raising revenues and diversifying financing sources for smart city development

Competing developmental priorities is commonplace in most developing countries, and this often leads to budget constraints among governments. To enable smart city development in developing countries, it is vital for governments to raise more revenue. Herein lies a paradigm shift regarding diversifying sources of financing and thinking about more innovative financing solutions for smart cities. While the appeal of multilateral developmental banks and foreign direct investment still holds in terms of socioeconomic and infrastructure development, governments should also think out of the box and consider using more viable and innovative mechanisms to raise funds. Besides land-based financing instruments, other innovative instruments that capitalise on the notion of 'congesters pay' can be rolled out to increase revenue for the government [35]. The 'metro plus property development approach' exemplified in the case of Shenzhen, China, also demonstrates a pragmatic approach in raising revenue for local government through mutually enhancing mechanisms that benefit the local governments, subway companies, and private real-estate developers [53].

(iii) Constructing regulatory frameworks for smart city governance

The introduction of IoT, big data, and other digital technology applications in the smart city indicate that users of technologies will be subject to different forms of technological risks such as data privacy and cybersecurity threats. Constructing basic regulatory frameworks for smart city governance is imperative, and governments in developing countries need to step up their efforts to chart clear blueprints for smart city development to minimise the risks and unintended consequences involved. To realise this vision, a nationwide governance framework for smart cities and legislations to govern the respective technology risks should be formulated. In developed countries, broad-stroke privacy laws (such as the General Data Protection Regulation [91], blockchain-based solutions [92], and education and training to increase public awareness, improve information literacy, and bridge the digital divide between the various sections of the populations [93] have been proposed to protect a citizen's data privacy and prevent cyberattacks. In some developing countries, privacy law has been enacted as a preliminary response to potential cyber-physical threats in smart city development [94]. Even though a policy framework for technology risk governance in developing countries remains nascent overall, case studies from China documenting how it streamlines governance processes in order to tap into the collective intelligence of the citizens could provide some inspiration for other developing countries in the formulation of their respective governance frameworks [52,95]. For instance, China has proposed a generic governance intelligence framework that can be directly incorporated into public services by tapping into the multiplier effects of citizen engagement and big data. By making citizens the central pillar of smart city governance, three layers of governance (the data-merging layer, the knowledge-discovery layer, and the decision-making layer) have been proposed to merge citizen-related data from multiple sources to build a panoramic dataset, map citizen profiles at both the individual and group levels, and use data mining to support decision-making [95]. Realising this objective will require governments to resolve the issue of fragmented authority and exert strong political will to ensure the effective enforcement of the smart city governance framework. To deal with volatility and policy uncertainty in governing novel technology in smart cities, governments in developing countries need to be strongly adaptive towards changing circumstances by maintaining the stability of governance structures and by being agile in contextualising policies that have worked in other countries and modifying them to suit local conditions [96,97].

(iv) Developing human capital and promoting digital inclusivity

Human capital is the most precious resource for any country and is one of the critical engines behind economic growth. This is all the more true for some developing countries that are currently experiencing a declining fertility rate but still have a substantial young working population [27]. Governments need to tap into this demographic dividend by developing human talent. Beyond providing basic education as an essential and universal investment in citizens, governments in developing countries need to craft longer-term visions for human capital strengthening by instituting various upskilling programmes, incentivising employers to provide regular training sessions, promoting knowledge and technology transfers from local and foreign private corporations, and encouraging stronger cultures of research and development within the higher institutions of learning. Alongside developing human capital, digital inclusivity is equally important. Governments need to be cognisant of the additional challenge presented by a large informal sector in most developing countries and ensure that the concerns of vulnerable populations are taken into consideration in the process of smart city development. Furthermore, financial incentives and other skills upgrading opportunities can be rolled out as measures to enable citizens who eke out a living in the informal economy to access higher education and other forms of learning opportunities.

(v) Creating a supportive ecosystem that nurtures start-ups and promotes public–private partnerships

A smart city is an ecosystem that comprises many different actors, subsystems, activity layers, and institutional logics, with their interactions augmented by various technological applications [15]. As such, governments need to create a supportive smart city sphere that nurtures all actors. Creating a more supportive and open business environment for start-ups, promoting pilots and trials for novel technology, and facilitating transparent and fair public–private partnerships in technology development are some of the measures that can be taken in novel technology adoption in the process of realising the technological visions of smart cities [98].

(vi) Encouraging citizen participation

The development of smart cities in developing countries should also incorporate the voices of citizens. Besides implementing public opinion polls, active citizen participation can be engaged through many technological solutions that incorporate e-government features in public service provisions and by utilising various social media outlets.

(vii) Promoting environmental sustainability

Finally, governments in developing countries should incorporate environmental sustainability as part of the policy narratives in smart city development [99]. While many eco-cities in the world advocate for this vision, the actual results of providing an environmentally friendly, comfortable urban life that is inclusive of all segments of the population remain ambiguous [34,45,48,80]. A recent study has concluded that the lack of a consolidated policy framework on the widescale utilisation of renewable energy is one of the biggest barriers to achieving renewable energy targets [100]. Instruments such as feed-in tariffs that are implemented with regulatory policies to facilitate new entrants into the renewable energy market and renewable portfolio standards whereby the government sets a quota for markets to produce and distribute a certain quantity of energies from renewable sources have been advocated to reduce carbon emissions and promote renewable energies in developed countries [100]. Different cities within a country can also be assessed and benchmarked for their capacity in achieving long-term sustainability [20].

**7. Conclusions**

Based on a systematic literature review, our study suggests that technology-enabled smart cities in developing countries can only be realised when concurrent socioeconomic, human, legal, and regulatory reforms are embedded in the long-term developmental trajectories of developing countries. Contextual conditions—including the state's social development, economic policy, and financial endowment; the technological literacy and willingness of citizens to partake in smart city development; and unique cultural factors—are important for smart city development in developing countries. Based on the insights gleaned from this review, some of the important future research directions include examining smart city policy diffusion and investigating how smart city lessons from one city are transferred and diffused domestically and internationally. In particular, more empirical case studies at either the country-level or city-level, focusing on understanding the implementation, design and governance of smart cities, are warranted. These case studies can be either in-depth single jurisdiction case studies or comparative case studies of multiple jurisdictions. These research endeavours will provide extensive contextualisations of governance lessons that are imperative in tackling the unique challenges faced by developing countries when rolling-out smart city initiatives.

conducted by the first author. Both authors contributed to the analysis and interpretation of the results, and the second author validated and edited the manuscript. Both authors have read and agreed to the published version of the manuscript.

**Funding:** This research is supported by the National Research Foundation, Prime Minister's Office, Singapore under its Campus for Research Excellence and Technological Enterprise (CREATE) programme.

**Acknowledgments:** Araz Taeihagh is grateful for the support provided by the National Research Foundation, Prime Minister's Office, Singapore under its Campus for Research Excellence and Technological Enterprise (CREATE) programme. Araz Taeihagh and Si Ying Tan appreciate Hazel Si Min Lim for reading the manuscript and providing suggestions.

**Conflicts of Interest:** The authors declare no conflicts of interest in this study.

**Appendix A: The Contexts of the 56 Studies Included in the Systematic Literature Review**

| | Study | Country (City)/Smart City Initiative | Methods/Study Design | Aims/Objectives of the Studies |
|---|---|---|---|---|
| 1 | A trivial approach for achieving smart city: a way forward towards a sustainable society [59] | Malaysia | Case study | This study discusses the 10 dimensions of a smart city using a case study from Malaysia |
| 2 | Smart city for development: towards a conceptual framework [54] | Indonesia | Case study | This study examines the three major goals of smart city development, discussing the resources and architectures needed in smart city development in Indonesia |
| 3 | The specificities and practical applications of Chinese eco-cities [50] | China | Case study | This study investigates the experiences of eco-city development in the Chinese context, specifying the definition and the goal of a smart city/eco-city |
| 4 | Sustainable smart city development framework for developing countries [32] | India | Case studies (three city-level cases) | This study discusses the four different dimensions of a smart city, using three city-level cases in India |
| 5 | The operationalizing aspects of smart cities: the case of Turkey's smart strategies [78] | Turkey | Case study | This study depicts some of the major instruments deployed to facilitate intelligent transport system in Turkey and the challenges involved |
| 6 | 'Actually existing sustainability' in urban China: National initiatives and local contestations [45] | China | Case study of major eco-city projects | This article describes the experiences of a few major eco-city developments in China, including the barriers/challenges involved |
| 7 | Smart cities in developing | India | Case study | This study discusses the dimensions of smart city |
| | economies: a literature review and policy insights [40] | | | development in India and the challenges involved in the implementation of smart cities |
| 8 | Effects of successful adoption of information technology enabled services in proposed smart cities of India: From user experience perspective [69] | India | Survey research (exploratory factor analysis) | This study describes smart city conceptualisations and goals, as well as the drivers that fuel smart city development in India |
| 9 | Alignment of IT authority and citizens of proposed smart cities in India: system security and privacy perspective [70] | India | Survey research (regression analysis) | This study describes the drivers of smart city development in India |
| 10 | Planning for sustainability in China's urban development: Status and challenges for Dongtan eco-city project [48] | China | Case study | This study depicts the implementation challenges of the Dongtan eco-city project in China |
| 11 | Smart city initiatives and the policy context: the case of the rapid business opening office in Mexico City [63] | Mexico | Case study | This study examines the case of smart city development in Mexico and depicts the key factors that drive smart city adoption |
| 12 | Smart cities: the main drivers for increasing the intelligence of cities [72] | Brazil | Comprehensive literature review and expert opinion survey | This study explores the definitions and concepts of 'smart city' for Brazil, and teases out the seven major drivers that characterise the intelligence of smart cities by engaging Brazilian experts from various fields |
| 13 | Low-carbon urban development strategy in Malaysia: the case of Iskandar Malaysia development corridor [60] | Malaysia | Case study (Iskandar development corridor) | This study illustrates several instruments applied in low-carbon city development in Malaysia |
| 14 | Smart city implementation framework for developing | Egypt | Case study | This study examines the policy process of smart city adoption in Egypt, including exploring the definition of smart city in the Egyptian context, its |
| | countries: the case of Egypt [81] | | | implementation challenges, and the domains and drivers of smart city development |
| 15 | The evolution of the smart cities agenda in India [41] | India | Case study (Discourse of Smart Cities Mission) | This study discusses several smart city instruments and depicts the challenges of smart city development in the Indian context |
| 16 | Prospect of Faridabad as a smart city: a review [42] | India | Case study | This study discusses the dimensions, instruments, and the drivers behind and barriers to smart city development in India |
| 17 | Smart city for development: a conceptual model for developing countries [83] | Developing countries as a whole | Review study | This study reviews the conceptualisation, drivers, and goals of smart city in developing countries as a whole |
| 18 | Smart neighbourhood: A TISM approach to reduce urban polarization for the sustainable development of smart cities [79] | India | Survey research | This study lays out and discusses the various components of a smart city based on the context of India |
| 19 | Can the smart city allure meet the challenges of Indian urbanization? [43] | India | Case study | This study examines the challenges of smart city adoption in India |
| 20 | Smart cities and the citizen-driven internet of things: a qualitative inquiry into an emerging smart city [44] | India (Hyderabad) | Case study (Hyderabad) | This study examines the goals and specific drivers behind and barriers to smart city adoption in Hyderabad, India |
| 21 | Critical success factors for eco-city development in China [71] | China | Survey research | This study illustrates the five major drivers behind smart cities in the context of China |
| 22 | Explaining the variety in smart eco city development in China: what policy network theory can teach us about overcoming barriers in implementation [49] | China (Shenzhen, Foshan, and Zhuhai cities) | Case studies | This study discusses the drivers and barriers/challenges facing smart city implementation by using three city-level cases in China |
| 23 | Smart city and quality of life: citizens' perception | Brazil (Curitiba City) | Questionnaire survey of 400 residents | This study illustrates the experience of smart city development in Brazil, |
| | in a Brazilian case study [3] | | | including the instruments deployed and the drivers behind smart city implementation |
| 24 | Platform ecosystems for Indonesia smart cities [55] | Indonesia | Case study | This study illustrates the dimensions and challenges of smart city adoption in Indonesia |
| 25 | Adoption of Internet of Things in India: a test of competing models using a structured equation modelling approach [67] | India | Survey research (structural equation modelling) | This study analyses the barriers to Internet-of things (IoT) use among consumers in India |
| 26 | Increasing collaboration and participation in smart city governance: a cross-case analysis of smart city initiatives [58] | Brazil (Rio de Janeiro, Porto Alegre, and Belo Horizonte) | Case studies | This study examines the implementation of smart cities (goals, instruments, drivers/enablers, and challenges) using three city-level cases in Brazil |
| 27 | A system view of smart mobility and its implications for Ghanaian cities [66] | Ghana | Case study | This study examines the experience of smart city adoption in Ghana, contemplating on its dimensions, goals, and challenges, including the problems that arise from the smart mobility initiative |
| 28 | Innovative civic engagement and digital urban infrastructure: lessons from 100 Smart Cities Mission in India [33] | India | Case study | This study describes the goals, barriers, and instruments of smart city adoption based on the context of the 100 Smart Cities Mission in India |
| 29 | Urban innovation through policy integration: critical perspectives from 100 smart cities mission in India [101] | India (Bhubaneswar) | Case study (Bhubaneswar city) | This study discusses the drivers behind and challenges facing smart city implementation/adoption in Bhubaneswar in India |
| 30 | Towards the right model of smart city governance in India [28] | India | Case study | This study examines the problems encountered in smart city implementation in India |
| 31 | Cutting through the clutter of smart city definitions: a reading into the smart city | India | Case studies (Lavasa, GIFT, New Town Kolkota, Jaipur) | This study examines smart city perceptions among the urban development professionals, and discusses the barriers/challenges |
| | perceptions in India [34] | | | involved in smart city development in India |
| 32 | The promise and performance of the world's first two zero-carbon eco-cities [80] | China | Case study | This study discusses the problems encountered in the development of the Dongtan eco-city in China |
| 33 | Barriers to the development of smart cities in Indian context [84] | India | Survey research | This study illustrates various barriers to smart city development based on a survey conducted in India |
| 34 | Developing smart cities: an integrated framework [83] | Developing countries as a whole | Review study | This study reviews the major dimensions, drivers, and challenges involved in smart city development in developing countries |
| 35 | Henry George and Mohring-Harwitz Theorems: lessons for financing smart cities in developing countries [35] | India | Case study | This study proposes three major strategies in financing smart cities in India |
| 36 | Smart funding options for developing smart cities: A proposal for India [36] | India | Case study | This study discusses means of financing smart cities in developing countries and the funding options for India |
| 37 | Promoting smart cities in developing countries: policy insights from Vietnam [73] | Vietnam | Survey research (500 experts, 10 cities in Vietnam) | This study examines the development and experiences of smart cities in Vietnam |
| 38 | Smart solutions shape for sustainable low-carbon future: a review on smart cities and industrial parks in China [46] | China | Country-level case study | This study examines the dimensions and drivers of smart cities using the context of industrial parks and low-carbon cities in China |
| 39 | Smart city with Chinese characteristics against the background of big data: idea, action, and risk [51] | China | Country-level case study | This study examines the conceptualisations, dimensions, goals, drivers, and challenges of smart city development in China |
| 40 | Developing a sustainable smart city framework for developing economies: an Indian context [68] | India | Empirical study (best worst method and interpretive structural | This study examines several enablers of smart city development and proposes smart city dimensions based on the context of India |

| | | | modelling approach) | |
|---|---|---|---|---|
| 41 | Towards a service-dominant platform for public value co-creation in a smart city: evidence from two metropolitan cities in China [52] | China (Shanghai and Guangzhou) | Case study | This study compares the experiences of rolling out two different models of service-dominant platforms in two major cities in China |
| 42 | Financing eco-cities and low-carbon cities: the case of Shenzhen International Low-Carbon City [53] | China (Shenzhen) | Case study | This paper discusses several financing options that Shenzhen has employed to finance the International Low-Carbon City project |
| 43 | Relevance of smart economy in smart cities in Africa [74] | Africa | Case study | This study discusses the implementation experiences and development of smart cities in the African context |
| 44 | Towards smart cities development: a study of public transport system and traffic-related air pollutants in Malaysia [61] | Malaysia | Case study | This study illustrates the implementation experience of one component of smart city—smart mobility—using the context of the smart city in Malaysia |
| 45 | Conceptualization to amendment: Kakinada as a smart city [37] | India (Kakinada) | Case study | This study describes the implementation experiences of smart city development in Kakinada, India |
| 46 | Smart city Nusantara Development through the application of Penta Helix Model [56] | Indonesia (Nusantara Development) | Case study | This study describes the implementation experiences of a smart city project (Nusantara Development) in Indonesia |
| 47 | The making of knowledge cities in Romania [64] | Romania | Case study | This study illustrates the application of technology as a driver for 'knowledge city' development in Romania |
| 48 | Smart sustainable city application: dimensions and developments [57] | Indonesia (Yogyakarta) | Case study | This study describes the implementation experiences of smart city development in Yogyakarta, Indonesia |
| 49 | Experimenting towards a low-carbon city: policy evolution and nested structure of innovation [47] | China (Shanghai) | Case study | This study describes the implementation experiences of low-carbon city development in Shanghai, China |
| 50 | Improving municipal solid waste collection | Nepal (Bharatpur | Case study | This study illustrates the experiences of municipal waste management in the context of |
|  | services in developing countries: a case of Bharatpur Metropolitan City, Nepal [65] | Metropolitan City) |  | smart city development in Nepal, and discusses the various challenges/barriers involved |
| 51 | Transportation planning aspects of a smart city: case study of GIFT City, Gujarat [38] | India (GIFT, Gujarat) | Case study | This study describes the experiences of smart transport development within the context of Gujarat International Finance Tec-City Company Limited (GIFTCL) in India |
| 52 | Design of IoT systems and analytics in the context of smart city initiatives in India [39] | India | Case study | This study describes the experiences of designing IoT systems in the smart city development context in India |
| 53 | Enabling technology for smart city transportation in developing countries [62] | Vietnam (Ho Chi Minh City) | Case study | This study examines the experiences of smart transportation in the context of Ho Chi Minh City, Vietnam |
| 54 | Achieving energy savings by intelligent transportation systems investments in the context of smart cities [77] | Developing countries as a whole | Review study | This study illustrates the three major characteristics of smart mobility (people-centric, data-driven, and powered by bottom-up initiatives) in the context of developing countries |
| 55 | Smart social development key for smart African cities [75] | African countries | Case study | This study discusses major regulatory/legal issues in smart city development in the context of satellite enhanced telemedicine and e-health in Africa |
| 56 | Transition to a low-carbon city: lessons learned from Suzhou in China [76] | China (Suzhou) | Case study | This study examines the implementation experiences of low-carbon city in Suzhou, China |

**References**

1. Yigitcanlar, T.; Kamruzzaman, M.; Buys, L.; Ioppolo, G.; Sabatini-Marques, J.; da Costa, E.M.; Yun, J.J. Understanding 'smart cities'': Intertwining development drivers with desired outcomes in a multidimensional framework.' *Cities* **2018**, *81*, 145–160.
2. Herrschel, T. Competitiveness AND sustainability: Can 'smart city regionalism' square the circle? *Urban Stud.* **2013**, *50*, 2332–2348.
3. Macke, J.; Casagrande, R.M.; Sarate, J.A.R.; Silva, K.A. Smart city and quality of life: Citizens' perception in a Brazilian case study. *J. Clean. Prod.* **2018**, *182*, 717–726.
4. Angelidou, M. Smart cities: A conjuncture of four forces. *Cities* **2015**, *47*, 95–106.
5. Winters, J. V Why are smart cotoes growing? Who moves who stays. *J. Reg. Sci.* **2011**, *51*, 253–270.
6. Allam, Z.; Dhunny, Z.A. On big data, artificial intelligence and smart cities. *Cities* **2019**, *89*, 80–91.
7. Bibri, S.E. The IoT for smart sustainable cities of the future: An analytical framework for sensor-based big data applications for environmental sustainability. *Sustain. Cities Soc.* **2018**, *38*, 230–253.

8. Bibri, S.E. On the sustainability of smart and smarter cities in the era of big data: an interdisciplinary and transdisciplinary literature review. *J. Big Data* **2019**, *6*, 25.
9. Martin, C.; Evans, J.; Karvonen, A.; Paskaleva, K.; Yang, D.; Linjordet, T. Smart-sustainability: A new urban fix? *Sustain. Cities Soc.* **2019**, *45*, 640–648.
10. Nesti, G. Defining and assessing the transformational nature of smart city governance: Insights from four European cases. *Int. Rev. Adm. Sci.* **2018**, 0020852318757063.
11. Razaghi, M.; Finger, M. Smart governance for smart cities. *Proc. IEEE* **2018**, *206*, 680–689.
12. Glasmeier, K.A.; Nebiolo, M. Thinking about smart cities: The travels of a policy idea that promises a great deal, but so far has delivered modest results. *Sustainability* 2016, *8*.
13. Glasmeier, A.; Christopherson, S. Thinking about smart cities. *Cambridge J. Reg. Econ. Soc.* **2015**, *8*, 3–12.
14. Castelnovo, W.; Misuraca, G.; Savoldelli, A. Smart cities governance: The need for a holistic approach to assessing urban participatory policy making. *Soc. Sci. Comput. Rev.* **2015**, *34*, 724–739.
15. Pierce, P.; Ricciardi, F.; Zardini, A. Smart Cities as organizational fields: A framework for mapping sustainability-enabling configurations. *Sustainability* 2017, *9*.
16. Meijer, A.; Bolívar, M.P.R. Governing the smart city: a review of the literature on smart urban governance. *Int. Rev. Adm. Sci.* **2015**, *82*, 391–408.
17. Lee, J.H.; Hancock, M.G.; Hu, M.-C. Towards an effective framework for building smart cities: Lessons from Seoul and San Francisco. *Technol. Forecast. Soc. Change* **2014**, *89*, 80–99.
18. Martin, C.J.; Evans, J.; Karvonen, A. Smart and sustainable? Five tensions in the visions and practices of the smart-sustainable city in Europe and North America. *Technol. Forecast. Soc. Change* **2018**, *133*, 269–278.
19. Akande, A.; Cabral, P.; Gomes, P.; Casteleyn, S. The Lisbon ranking for smart sustainable cities in Europe. *Sustain. Cities Soc.* **2019**, *44*, 475–487.
20. Shmelev, S.E.; Shmeleva, I.A. Multidimensional sustainability benchmarking for smart megacities. *Cities* **2019**, *91*, 134–163.
21. Silva, B.N.; Khan, M.; Han, K. Towards sustainable smart cities: A review of trends, architectures, components, and open challenges in smart cities. *Sustain. Cities Soc.* **2018**, *38*, 697–713.
22. Habibzadeh, H.; Nussbaum, B.H.; Anjomshoa, F.; Kantarci, B.; Soyata, T. A survey on cybersecurity, data privacy, and policy issues in cyber-physical system deployments in smart cities. *Sustain. Cities Soc.* **2019**, *50*, 101660.
23. Ruhlandt, R.W.S. The governance of smart cities: A systematic literature review. *Cities* **2018**, *81*, 1–23.
24. Neirotti, P.; De Marco, A.; Cagliano, A.C.; Mangano, G.; Scorrano, F. Current trends in smart city initiatives: Some stylised facts. *Cities* **2014**, *38*, 25–36.
25. Yigitcanlar, T.; Han, H.; Kamruzzaman, M.; Ioppolo, G.; Sabatini-Marques, J. The making of smart cities: Are Songdo, Masdar, Amsterdam, San Francisco and Brisbane the best we could build? *Land use policy* **2019**.
26. Smith, K. The inconvenient truth about smart cities 2017.
27. United Nations *World urbanization prospects: The 2018 revision [key facts]*; 2018;
28. Praharaj, S.; Han, J.H.; Haken, S. Towards the right model of smart city governance in India. *Int. J. Sustain. Dev. Plan.* **2018**, *13*, 171–186.
29. Chandrasekar, K.S.; Bajracharya, Bhishna O'Hare, D. A comparative analysis of smart city initiatives by China and India - Lessons for India. In Proceedings of the 9th International Urban Design conference: Smart cities for 21st century Australia: How urban design innovation can change our cities.; Canberra, 2016.
30. Thomas, J.; Harden, A. Methods for the thematic synthesis of qualitative research in systematic reviews. *BMC Med. Res. Methodol.* **2008**, *8*, 45.
31. Miles, M.B.; Huberman, A.M. *Qualitative data analysis: An expanded sourcebook*; 2nd editio.; SAGE Publications: Thousand Oaks, CA, 1994;
32. Bhattacharya, T.R.; Bhattacharya, A.; Mclellan, B.; Tezuka, T. Sustainable smart city development framework for developing countries. *Urban Res. Pract.* **2018**, 1–33.
33. Praharaj, S.; Han, J.H.; Hawken, S. Innovative civic engagement and digital urban infrastructure: Lessons from 100 smart cities mission in India. *Procedia Eng.* **2017**, *180*, 1423–1432.
34. Praharaj, S.; Han, H. Cutting through the clutter of smart city definitions: A reading into the smart city perceptions in India. *City, Cult. Soc.* **2019**.
35. Mishra, A.K. Henry George and Mohring–Harwitz Theorems: Lessons for financing smart cities in developing countries. *Environ. Urban. Asia* **2019**, *10*, 13–30.
36. Vadgama, C.V.; Khutwad, A.; Damle, M.; Patil, S. Smart funding options for developing smart cities: A proposal for India. *Indian J. Sci. Technol. Vol. 8, Issue 34, December 2015* **2015**.

37. Chintagunta, L.; Raj, P.; Narayanaswami, S. Conceptualization to amendment: Kakinada as a smart city. *J. Public Aff.* **2019**, *19*, e1879.
38. Reddy, D.S.; Babu, K.V.G.; Murthy, D.L.N. Transportation planning aspects of a smart city–case study of GIFT City, Gujarat. *Transp. Res. Procedia* **2016**, *17*, 134–144.
39. Vijai, P.; Sivakumar, P.B. Design of IoT systems and analytics in the context of smart city initiatives in India. *Procedia Comput. Sci.* **2016**, *91*, 583–588.
40. Chatterjee, S.; Kar, A.K. Smart Cities in developing economies: A literature review and policy insights. In Proceedings of the 2015 International Conference on Advances in Computing, Communications and Informatics (ICACCI); 2015; pp. 2335–2340.
41. Hoelscher, K. The evolution of the smart cities agenda in India. *Int. Area Stud. Rev.* **2016**, *19*, 28–44.
42. Jamal, S.; Sen, A. Prospect of Faridabad as a smart city: A review. In *Making cities resilient*; Sharma, V.R., Chandrakanta, Eds.; Springer International Publishing: Cham, Switzerland, 2019; pp. 39–52.
43. Kumar, A. Can the smart city allure meet the challenges of Indian urbanization? In *Sustainable smart cities in India: Challenges and future prespectives*; Sharma, P., Rajput, S., Eds.; Springer International Publishing: Cham, Switzerland, 2017; pp. 17–40.
44. Kummitha, R.K.R.; Crutzen, N. Smart cities and the citizen-driven internet of things: A qualitative inquiry into an emerging smart city. *Technol. Forecast. Soc. Change* **2019**, *140*, 44–53.
45. Chang, I.-C.C. "Actually existing sustainabilities" in urban China: National initiatives and local contestations. *Sustainability* **2018**, *11*, 216–228.
46. Wang, Y.; Ren, H.; Dong, L.; Park, H.-S.; Zhang, Y.; Xu, Y. Smart solutions shape for sustainable low-carbon future: A review on smart cities and industrial parks in China. *Technol. Forecast. Soc. Change* **2019**, *144*, 103–117.
47. Peng, Y.; Bai, X. Experimenting towards a low-carbon city: Policy evolution and nested structure of innovation. *J. Clean. Prod.* **2018**, *174*, 201–212.
48. Cheng, H.; Hu, Y. Planning for sustainability in China's urban development: Status and challenges for Dongtan eco-city project. *J. Environ. Monit.* **2010**, *12*, 119–126.
49. Lu, H.; de Jong, M.; ten Heuvelhof, E. Explaining the variety in smart eco city development in China-What policy network theory can teach us about overcoming barriers in implementation? *J. Clean. Prod.* **2018**, *196*, 135–149.
50. Bao, S.; Toivonen, M. The specificities and practical applications of Chinese eco-cities. *J. Sci. Technol. Policy Manag.* **2014**, *5*, 162–176.
51. Wu, Y.; Zhang, W.; Shen, J.; Mo, Z.; Peng, Y. Smart city with Chinese characteristics against the background of big data: Idea, action and risk. *J. Clean. Prod.* **2018**, *173*, 60–66.
52. Yu, J.; Wen, Y.; Jin, J.; Zhang, Y. Towards a service-dominant platform for public value co-creation in a smart city: Evidence from two metropolitan cities in China. *Technol. Forecast. Soc. Change* **2019**, *142*, 168–182.
53. Zhan, C.; de Jong, M. Financing eco cities and low carbon cities: The case of Shenzhen International Low Carbon City. *J. Clean. Prod.* **2018**, *180*, 116–125.
54. Achmad, K.A.; Nugroho, L.E.; Djunaedi, A.; Widyawan Smart city for development: Towards a conceptual framework. In Proceedings of the 2018 4th International Conference on Science and Technology (ICST); 2018; pp. 1–6.
55. Mahesa, R.; Yudoko, G.; Anggoro, Y. Platform ecosystems for Indonesia smart cities. In Proceedings of the 2018 International Conference on Computer, Control, Informatics and its Applications (IC3INA); 2018; pp. 34–39.
56. Effendi, D.; Syukri, F.; Subiyanto, A.F.; Utdityasan, R.N. Smart city Nusantara development through the application of Penta Helix model (A practical study to develop smart city based on local wisdom). In Proceedings of the 2016 International Conference on ICT For Smart Society (ICISS); 2016; pp. 80–85.
57. Larasati, N.; Handayaningsih, S.; Sumarsono, S. Smart sustainable city application: Dimensions and developments. In Proceedings of the 2018 IEEE 3rd International Conference on Communication and Information Systems (ICCIS); 2018; pp. 122–126.
58. Viale Pereira, G.; Cunha, M.A.; Lampoltshammer, T.J.; Parycek, P.; Testa, M.G. Increasing collaboration and participation in smart city governance: a cross-case analysis of smart city initiatives. *Inf. Technol. Dev.* **2017**, *23*, 526–553.
59. Jnr, B.A.; Majid, M.A.; Romli, A. A trivial approach for achieving smart city: A way forward towards a sustainable society. In Proceedings of the 2018 21st Saudi Computer Society National Computer Conference (NCC); 2018; pp. 1–6.
60. Ho, C.S.; Matsuoka, Y.; Simson, J.; Gomi, K. Low carbon urban development strategy in Malaysia- The case of Iskandar Malaysia development corridor. *Habitat Int.* **2013**, *37*, 43–51.

61. Brohi, S.N.; Pillai, T.R.; Asirvatham, D.; Ludlow, D.; Bushell, J. Towards smart cities development: A study of public transport system and traffic-related air pollutants in Malaysia. *IOP Conf. Ser. Earth Environ. Sci.* **2018**, *167*, 12015.
62. Zan, T.T.T.; Gueta, L.B.; Okochi, T. Enabling technology for smart city transportation in developing countries. In Proceedings of the 2015 IEEE International Conference on Smart City/SocialCom/SustainCom (SmartCity); 2015; pp. 170–174.
63. Gil-Garcia, J.R.; Aldama-Nalda, A. Smart city initiatives and the policy context: The case of the rapid business opening office in Mexico City. In Proceedings of the ICEGOV '13; Seoul, 2013.
64. Elena, C. The making of Knowledge Cities in Romania. *Procedia Econ. Financ.* **2015**, *32*, 534–541.
65. Rai, K.R.; Nepal, M.; Khadayat, S.M.; Bhardwaj, B. Improving municipal solid waste collection services in developing countries: A case of Bharatpur metropolitan city, Nepal. *Sustainability* 2019, *11*.
66. Peprah, C.; Amponsah, O.; Oduro, C. A system view of smart mobility and its implications for Ghanaian cities. *Sustain. Cities Soc.* **2019**, *44*, 739–747.
67. Mital, M.; Chang, V.; Choudhary, P.; Papa, A.; Pani, A.K. Adoption of Internet of Things in India: A test of competing models using a structured equation modeling approach. *Technol. Forecast. Soc. Change* **2018**, *136*, 339–346.
68. Yadav, G.; Mangla, S.K.; Luthra, S.; Rai, D.P. Developing a sustainable smart city framework for developing economies: An Indian context. *Sustain. Cities Soc.* **2019**, *47*, 101462.
69. Chatterjee, S.; Kar, A.K. Effects of successful adoption of information technology enabled services in proposed smart cities of India. *J. Sci. Technol. Policy Manag.* **2017**, *9*, 189–209.
70. Chatterjee, S.; Kar, A.K.; Gupta, M.P. Alignment of IT authority and citizens of proposed smart cities in India: System security and privacy perspective. *Glob. J. Flex. Syst. Manag.* **2018**, *19*, 95–107.
71. Liu, J.; Low, S.P.; Wang, L.F. Critical success factors for eco-city development in China. *Int. J. Constr. Manag.* **2018**, *18*, 497–506.
72. Azevedo Guedes, L.A.; Carvalho Alvarenga, J.; Dos Santos Sgarbi Goulart, M.; Rodriguez y Rodriguez, V.M.; Pereira Soares, A.C. Smart cities: The main drivers for increasing the intelligence of cities. *Sustainability* 2018, *10*.
73. Vu, K.; Hartley, K. Promoting smart cities in developing countries: Policy insights from Vietnam. *Telecomm. Policy* **2018**, *42*, 845–859.
74. Mboup, G.; Oyelaran-Oyeyinka, B. Relevance of smart economy in smart cities in Africa. In *Smart economy in smart African cities: Sustainable, inclusive, resilient and prosperous*; Mboup, G., Oyelaran-Oyeyinka, B., Eds.; Springer Nature Singapore: Singapore, 2018; pp. 1–50.
75. Idele, P.; Mboup, G. Smart social development key for smart African cities. In *Smart economy in smart African cities: Sustainable, inclusive, resilient and prosperous*; Mboup, G., Oyelaran-Oyeyinka, B., Eds.; Springer Nature Singapore: Singapore, 2018; pp. 393–422.
76. Liu, W.; Wang, C.; Xie, X.; Mol, A.P.J.; Chen, J. Transition to a low-carbon city: lessons learned from Suzhou in China. *Front. Environ. Sci. Eng.* **2012**, *6*, 373–386.
77. Chen, Y.; Ardila-Gomez, A.; Frame, G. Achieving energy savings by intelligent transportation systems investments in the context of smart cities. *Transp. Res. Part D Transp. Environ.* **2017**, *54*, 381–396.
78. Tekin Bilbil, E. The operationalizing aspects of smart cities: The case of Turkey's smart strategies. *J. Knowl. Econ.* **2017**, *8*, 1032–1048.
79. Kumar, H.; Singh, M.K.; Gupta, M.P.; Madaan, J. Smart neighbourhood: A TISM approach to reduce urban polarization for the sustainable development of smart cities. *J. Sci. Technol. Policy Manag.* **2018**, *9*, 210–226.
80. Premalatha, M.; Tauseef, S.M.; Abbasi, T.; Abbasi, S.A. The promise and the performance of the world's first two zero carbon eco-cities. *Renew. Sustain. Energy Rev.* **2013**, *25*, 660–669.
81. Hamza, K. Smart city implementation framework for developing countries: The case of Egypt. In *Smarter as the new urban agenda: A comprehensive view of the 21st century city*; Gil-Garcia, J.R., Pardo, T.A., Nam, T., Eds.; Springer International Publishing Switzerland: Cham, Heidelberg, New York, Dordrecht, London, 2016; pp. 171–190.
82. Joia, L.A.; Kuhl, A. Smart city for development: A conceptual model for developing countries. In Proceedings of the 15th IFIP WG 9.4 International Conference on Social Implications of Computers in Developing Countries, ICT4D 2019; Dar es Salaam, Tanzania, 2019; pp. 114–203.
83. Joshi, S.; Saxena, S.; Godbole, T.; Shreya Developing smart cities: An integrated framework. *Procedia Comput. Sci.* **2016**, *93*, 902–909.
84. Rana, N.P.; Luthra, S.; Mangla, S.K.; Islam, R.; Roderick, S.; Dwivedi, Y.K. Barriers to the development of smart cities in Indian context. *Inf. Syst. Front.* **2019**, *21*, 503–525.
85. UNDP *Human Development Report 2019 Beyond income, beyond averages, beyond today: Inequalities in human*

*development in the 21st century*; 2019;

86. Taeihagh, A. Crowdsourcing, sharing economies and development. *J. Dev. Soc.* **2017**, *33*, 191–222.
87. Biswas, K.; Muthukkumarasamy, V. Securing Smart Cities Using Blockchain Technology. In Proceedings of the 2016 IEEE 18th International Conference on High Performance Computing and Communications; IEEE 14th International Conference on Smart City; IEEE 2nd International Conference on Data Science and Systems (HPCC/SmartCity/DSS); 2016; pp. 1392–1393.
88. Cain Piers, K.B. Information, Not Technology, Is Essential to Accountability: Electronic Records and Public-Sector Financial Management. *Inf. Soc.* **2001**, *17*, 247–258.
89. Ramaswami, A.; Russell, A.G.; Culligan, P.J.; Sharma, K.R.; Kumar, E. Meta-principles for developing smart, sustainable, and healthy cities. *Science (80-. ).* **2016**, *352*, 940 LP-- 943.
90. World Health Organization *Copenhagen Consensus of Mayors: Healthier and happier cities for all a transformative approach for safe, inclusive, sustainable and resilient societies*; 2018;
91. Wachter, S. The GDPR and the Internet of Things: a three-step transparency model. *Law, Innov. Technol.* **2018**, *10*, 266–294.
92. Makhdoom, I.; Zhou, I.; Abolhasan, M.; Lipman, J.; Ni, W. PrivySharing: A blockchain-based framework for privacy-preserving and secure data sharing in smart cities. *Comput. Secur.* **2020**, *88*, 101653.
93. Lam, P.T.I.; Ma, R. Potential pitfalls in the development of smart cities and mitigation measures: An exploratory study. *Cities* **2019**, *91*, 146–156.
94. Ni Loideain, N. Cape Town as a smart and safe city: implications for governance and data privacy. *Int. Data Priv. Law* **2017**, *7*, 314–334.
95. Ju, J.; Liu, L.; Feng, Y. Citizen-centered big data analysis-driven governance intelligence framework for smart cities. *Telecomm. Policy* **2018**, *42*, 881–896.
96. Janssen, M.; van der Voort, H. Adaptive governance: Towards a stable, accountable and responsive government. *Gov. Inf. Q.* **2016**, *33*, 1–5.
97. Soe, R.-M.; Drechsler, W. Agile local governments: Experimentation before implementation. *Gov. Inf. Q.* **2018**, *35*, 323–335.
98. Tan, S.Y.; Taeihagh, A.; Tripathi, A. Tensions and antagonistic interactions of technological risks and ethical concerns of robotics and autonomous systems in long-term care. *Forthcoming* **2019**.
99. Yigitcanlar, T.; Kamruzzaman, M.; Foth, M.; Sabatini-Marques, J.; da Costa, E.; Ioppolo, G. Can cities become smart without being sustainable? A systematic review of the literature. *Sustain. Cities Soc.* **2019**, *45*, 348–365.
100. Malik, K.; Rahman, S.M.; Khondaker, A.N.; Abubakar, I.R.; Aina, Y.A.; Hasan, M.A. Renewable energy utilization to promote sustainability in GCC countries: policies, drivers, and barriers. *Environ. Sci. Pollut. Res.* **2019**, *26*, 20798–20814.
101. Praharaj, S.; Han, J.H.; Hawken, S. Urban innovation through policy integration: Critical perspectives from 100 smart cities mission in India. *City, Cult. Soc.* **2018**, *12*, 35–43.